\documentclass[journal=acsnano,manuscript=article]{achemso}

\usepackage[version=3]{mhchem} 
\usepackage{xcolor}
\usepackage[normalem]{ulem}

\usepackage{graphicx}
\usepackage{xcolor}
\usepackage{amsmath}
\usepackage{amssymb}
\usepackage{amsfonts}
\usepackage{dsfont}
\usepackage{diagbox}
\usepackage{upgreek}
\usepackage{hyperref}
\usepackage{siunitx}

\author{Erik Olsén}
\affiliation{\small Department of Physics, Chalmers University of Technology, Gothenburg, Sweden}
\email{olsene@chalmers.se}
\author{Benjamin Midtvedt}
\affiliation{{\small Department of Physics, University of Gothenburg, Gothenburg, Sweden}}
\author{Adrián González}
\author{Fredrik Eklund}
\affiliation{\small Department of Physics, Chalmers University of Technology, Gothenburg, Sweden}
\author{Katarzyna Ranoszek-Soliwoda}
\affiliation{{\small Department of Materials Technology and Chemistry, Faculty of Chemistry, University of Łódź,  Łódź, Poland}}
\author{Jaroslaw Grobelny}
\affiliation{{\small Department of Materials Technology and Chemistry, Faculty of Chemistry, University of Łódź,  Łódź, Poland}}
\author{Giovanni Volpe}
\affiliation{{\small Department of Physics, University of Gothenburg, Gothenburg, Sweden}}
\author{Malgorzata Krzyzowska}
\affiliation{{\small Department of Rheumatology and Inflammation Research, Sahlgrenska Academy, University of Gothenburg, Gothenburg, Sweden}}
\alsoaffiliation{{\small Department of Nanobiology and Biomaterials, Military Institute of Hygiene and Epidemiology, Warsaw, Poland}}
\author{Fredrik H{\"o}{\"o}k}
\affiliation{\small Department of Physics, Chalmers University of Technology, Gothenburg, Sweden}
\email{fredrik.hook@chalmers.se}
\author{Daniel Midtvedt}
\affiliation{{\small Department of Physics, University of Gothenburg, Gothenburg, Sweden}}
\email{daniel.midtvedt@physics.gu.se}

\title{Label-free optical quantification of material composition of suspended virus-gold nanoparticle complexes}

\begin{document}


\begin{abstract}
\noindent 
The interaction between metallic and biological nanoparticles (NPs) is widely used in various biotechnology and biomedical applications. However, detailed characterization of this type of interaction is challenging due to a lack of high-throughput techniques that can quantify both size and composition of suspended NP complexes. 
Here, we introduce a technique called ``twilight nanoparticle tracking analysis'' (tNTA) and demonstrate that it provides a quantitative relationship between the measured optical signal and the composition of suspended dielectric-metal NP complexes.
We assess the performance of tNTA by analyzing the selective binding of tannic acid-modified gold nanoparticles (TaAuNPs) to herpes simplex viruses (HSV). 
Our results show that TaAuNPs bind specifically to HSV without causing substantial changes in the size or refractive index of the virus, suggesting that the binding does not cause virus disruption. 
Instead, the anti-viral properties of TaAuNPs appear to stem from direct particle binding to the virus.

\end{abstract}


\section{Introduction}

\noindent Biological nanoparticles (NPs) are a diverse category that includes everything from intra- and extracellular vesicles used in cellular signalling processes\cite{andaloussi2013extracellular} to pathogenic viruses such as hepatitis\cite{fanning2019therapeutic}, corona \cite{cui2019origin} and herpes\cite{connolly2021structural}. 
Biological NPs have also served as inspiration for the supramolecular design of next-generation nanoscopic drug carriers\cite{hou2021lipid}.
Because of their faint optical contrasts, these biological entities are often identified using functionalised metallic NPs, which increase the contrast in both single-particle imaging\cite{manzo2015review} and diagnostic applications\cite{farka2017nanoparticle}. 
Functionalised metallic NPs are also commonly used in drug delivery and local treatment\cite{yang2019gold,mitchell2021engineering}, as well as to specifically inhibit the activity of biological pathogens, including bacteria\cite{ahmed2016future,yu2016inhibition} and viruses\cite{papp2010inhibition,orlowski2014tannic,cagno2018broad,tang2020materials}. 
As a specific example, tannic-acid-modified silver NPs have been used to inhibit herpes simplex-2 virus (HSV-2) infections in both cell and animal models\cite{orlowski2014tannic,orlowski2018tannic}.
However, the detailed nature of the interaction between tannic-acid-modified metal NPs and HSV-2 as well as the underlying inhibitory mechanisms remain unclear, largely due to challenges related to non-perturbing experimental characterisation of the resulting virus/gold NP complexes.
Detailed characterisation of such particle complexes is typically obtained using cryogenic transmission electron microscopy (CryoTEM)\cite{papp2010inhibition,cagno2018broad,zhigaltsev2022synthesis}, which can provide nanometre resolution snapshot information, but is an ex situ analysis technique with low throughput, which limits its applicability for investigating biological systems.

Although label-free optical microscopy is a common technique for high-throughput in-situ characterisation of single-NPs \cite{priest2021scattering,kashkanova2021precision}, its ability to characterise NP complexes is currently limited. 
This limitation arises from the challenge of achieving sufficient sensitivity while also obtaining quantitative, material-specific signals in multi-component systems. 
Typically, single NP characterisation is based on relating the NP diffusivity to size\cite{van2010optical} and/or its optical scattering signal to size and refractive index (RI)\cite{lee2007characterizing,khadir2020full,midtvedt2021fast}. 
Although the scattering intensity is not a material-specific signal\cite{priest2021scattering}, the complex-valued scattered field, which can be quantified through quantitative phase microscopy methods such as holographic imaging, does provide material-specific information through the real and imaginary parts of the complex polarizability\cite{khadir2020full,nguyen2023label}.
This complex-valued scattered signal has been used to distinguish different types of particles from each other based on the real and imaginary parts\cite{khadir2020full,nguyen2023label}, but so far the investigated particles have been limited to single material NPs.
A specific complication related to quantification of the complex-valued scattered signal from individual suspended NPs using quantitative phase imaging is the low signal-to-background ratio of such techniques compared to dark-field techniques.

Recent advances in optical interferometric imaging have made it possible to detect objects as small as lipid vesicles and single molecules by controlling the signal-to-background ratio\cite{priest2021scattering,kashkanova2021precision,vspavckova2022label}.
Two examples of interferometric techniques with controllable signal-to-background ratios for the detection of optically very faint particles are interferometric scattering microscopy (iSCAT)\cite{priest2021scattering,kashkanova2021precision} and nanochannel scattering microscopy (NSM)\cite{vspavckova2022label}, but the depth dependence of the iSCAT signal\cite{dong2021fundamental} and uncertain-confined diffusivity-size relations in NSM\cite{bian2016111,olsen2021diffusion} make it difficult to use these configurations to relate particle measurements to properties such as the complex polarizability or RI.
A third implementation, which combines a transmission geometry with a controllable signal-to-background ratio, can be achieved by using a low-frequency attenuation filter (LFAF)\cite{pham2012digital}. 
The LFAF is a semitransparent disk in the back-focal plane of the objective (or a conjugate to the back-focal plane) that reduces the intensity of the unscattered light while leaving the scattered light unaffected, thereby enhancing the signal-to-noise ratio. 
This approach has enabled distinguishing metallic from dielectric NPs\cite{liebel2021widefield,saemisch2021one}.
However, the LFAF disturbs the relation between the background and the interferometric particle signal. For this reason, quantitative measurements of NP complexes containing both metallic and dielectric NPs have not been demonstrated.

Here, we introduce a new technique called ``twilight nanoparticle tracking analysis'' (tNTA), which can quantify the size and the absolute amounts of dielectric and metallic material in suspended NP complexes. 
We achieve a four-fold improvement in the detection limit without loss of quantitative scattering information by taking advantage of the increase in signal-to-noise ratio obtained by employing an LFAF. 
We quantify and correct the effect of the LFAF by calibrating the system on monodisperse nanoparticles with known properties. 
Additionally, by tracking NP complexes under flow, we simultaneously determine the hydrodynamical radius, RI, number of bound metallic NPs, and dry mass of individual NP complexes.
To demonstrate the capabilities of tNTA, we quantify the specific NP interaction between suspended tannic-acid-modified gold NPs (TaAuNPs) and HSV-2. 
The information obtained from tNTA regarding the number of bound TaAuNPs per HSV-2, the hydrodynamic radius, and the RI of individual HSV-2/TaAuNP complexes, suggests that physical hindrance is the primary anti-viral pathway for TaAuNPs rather than TaAuNP-induced virus rupture.

\section*{Results and Discussion}
\subsection*{Operating principle of tNTA}

\begin{figure}
\includegraphics[width=1\textwidth]{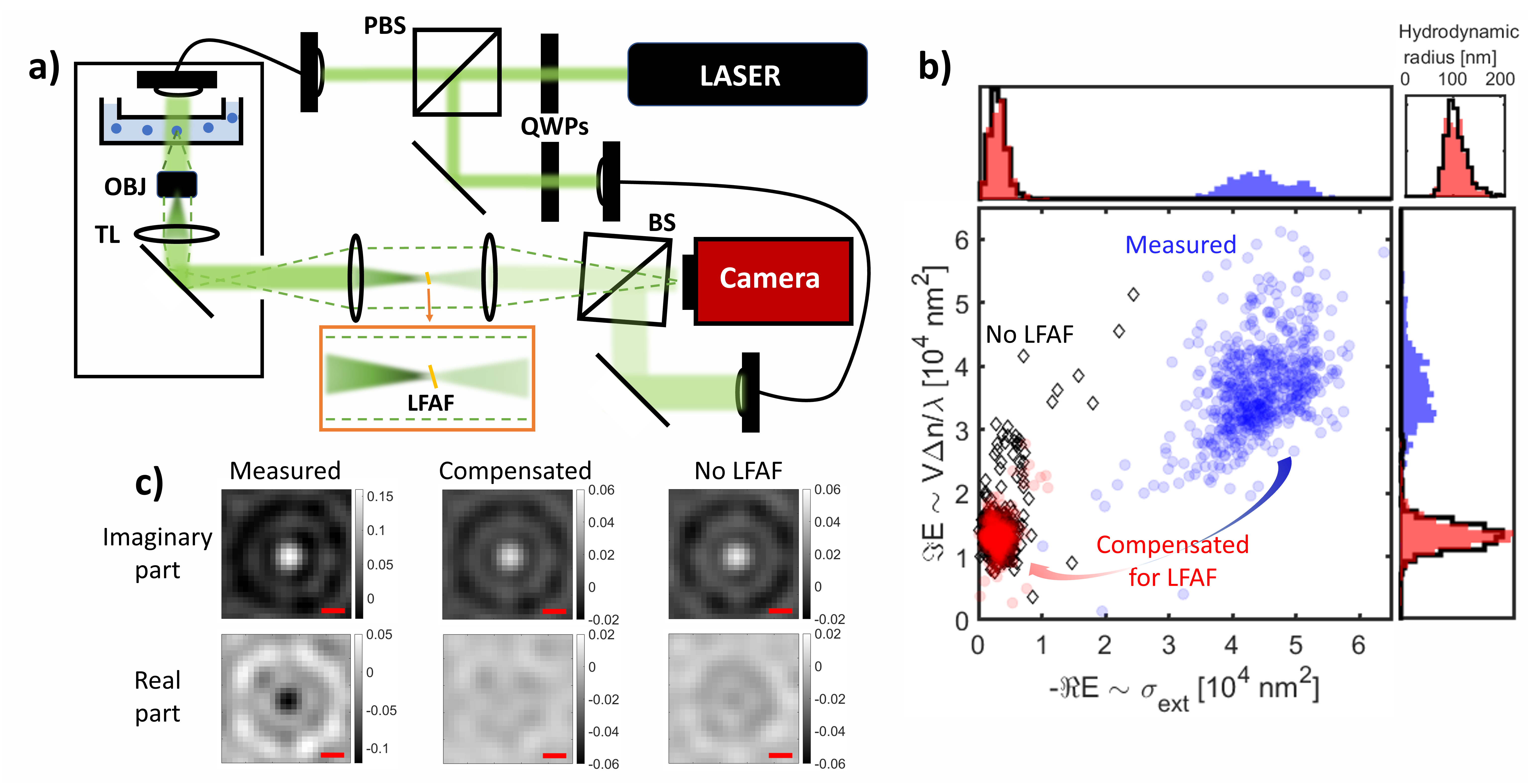} 
\caption{{\bf Schematic illustration of a twilight off-axis holographic microscope and measurement principle.} 
{\bf a}, The laser beam is split into two beams by the polarising beam splitter (PBS): one beam goes through the sample, while the other beam acts as a reference. The low frequency attenuation filter (LFAF) reduces the amplitude of the unscattered light of the sample beam while having a negligible effect on the particle signal. The reduction of the unscattered light is highlighted in the zoomed-in inset. The LFAF is slightly tilted in order to direct the reflected light away from the optical axis. BS: beam splitter, OBJ: objective, TL: tube lens, and HWPs: Half-wave plates. 
{\bf b}, The calibration of the LFAF compensation factor due to the background normalisation is performed by first quantifying the optical signal from \SI{105}{\nm} radius polystyrene spheres both with (blue symbols and blue histograms) and without (open, black symbols and black outlined histograms) the LFAF. The compensation factor is determined by requiring that the median particle signal after compensation should be the same for the two measurements. The red symbols and histograms correspond to the population measured using the LFAF after compensation. Note that the inferred optical signal and the hydrodynamical size are similar after compensation. 
{\bf c}, Images of the real and imaginary parts of the optical signal both with and without the LFAF, showing the similarity of the optical signal after compensation and without the LFAF. The scalebars correspond to \SI{500}{\nm}.}
\label{fig:1}
\end{figure}

Off-axis holography is an imaging technique where the complex-valued optical signal, $E_{s}$, is quantified via the interference with an external reference beam that hits the camera at a small angle relative to the sample beam\cite{kim2010principles} (Methods, ''Data analysis''). 
The twilight off-axis holographic microscope presented in this paper is an off-axis holographic microscope equipped with a LFAF. 
In our setup, the LFAF is a gold disc with a thickness and diameter of \SI{55}{\nano\meter} and $\sim$\SI{500}{\micro\meter}, respectively, located at the intermediate focal plane of a 4f system (Figure \ref{fig:1}a). 

The purpose of the LFAF is to selectively attenuate the unscattered light while leaving the scattered light from the particles unchanged, which is achieved by utilising that the nanoparticle scattering mostly occur at angles different from that of the illumination in the case of an incoming plane wave.
The signal-to-background ratio is thereby increased, while the coherence between scattered and unscattered field is maintained, enabling quantification of the optical field scattered from the particles (Supplementary information, Figure S1))\cite{pham2012digital}. 

The optical signal from the sample is the sum of the unscattered background $E_0$ and particle signal $E_{\rm p}$. Since the LFAF selectively attenuates the unscattered light, the optical signal at the camera becomes
\begin{equation}
E_{s}= aE_0 \left(1+ (E_{\rm p}/aE_0)\right),
\label{Eq:2}
\end{equation}
where $|a|<1$ is a complex constant that encodes the effect of the LFAF, and depends on the geometry of the LFAF. 
In the absence of the LFAF, $a\equiv 1$. 
From this relation, it is clear that the LFAF perturbs the interferometric signal $\left(1+ (E_{\rm p}/aE_0)\right)$ that is measured in off-axis holography.  
In order to quantify the scattered field $E_{\rm p}$ from the particle, it is therefore necessary to determine the parameter $a$.

To calibrate the parameter $a$, we measured a monodisperse sample with known optical properties with and without the LFAF. 
Specifically, we measured \SI{105}{\nm} radius polystyrene spheres suspended in a microfluidic channel under flow  (Methods, ''Experimental conditions'') (Figure \ref{fig:1}b). 
By requiring that the median particle signal is the same both with and without the LFAF, we find $a=(0.251 \pm 0.004) \times{\rm exp}(-0.59i\pm 0.01i)$ (Supplementary information, Section 1). 
As the presence of the LFAF increases the noise level with 15-30\% (Supplementary information, Figure S3), this value of $a$ implies an approximately four-fold improvement of the smallest detectable optical mass compared to ordinary off-axis holography.

Our observations support the assumption that the LFAF does not influence the scattered field from the particles, and that the effect of the LFAF can be numerically compensated for in the post-analysis. 
Specifically, the observed hydrodynamic radius distribution is not affected by the presence of the LFAF, indicating that the performance of the tracking algorithm is not influenced by the LFAF (Supplementary information, Section 2 and Supplementary Figure S4). 
Moreover, using the obtained correction factor, the tNTA particle images become very similar to images obtained when the LFAF is absent (Figure \ref{fig:1}c), and the optical signal displays no depth-dependence across the channel (Supplementary information, Figure S2).


For sub-wavelength-sized particles, material specificity is contained in the relation between the integrated real and imaginary parts as they relate to different physical aspects of light--matter interaction. 
Specifically, if the image is normalised such that $E_0=1$, the integrated imaginary part, $\Im \text{E}$, relates to the phase shift as
\begin{equation}
    \Im \text{E}=\iint \text{Im}(E_{\rm p}/E_0)\; dA\approx \iint \phi(1+E_{\rm p}/E_0)\; dA = \frac{2\pi}{\lambda} V\Delta n,
    \label{Eq:phase}
\end{equation}
whereas the integrated real part, $\Re \text{E}$, relates to the extinction cross section, $\sigma$, as (Supplementary Information, Section 3)\cite{khadir2020full}
\begin{equation}
    \Re \text{E}=\iint \text{Re}(E_{\rm p}/E_0)\; dA \approx -\sigma/2.
    \label{Eq:realpart}
\end{equation}
Thus, the values of $\Im \text{E}$ and $\Re \text{E}$ can be directly related to material-specific optical properties of NPs: $\Im \text{E}$ relates to the integrated phase shift, while $\Re \text{E}$ relates to the light extinction (e.g., due to light absorption within the NP).
This permits one to distinguish, for example, dielectric particles (which primarily induce a light phase shift) and small metallic NPs (which absorb light to a larger degree)\cite{khadir2020full}.
Since polystyrene spheres are dielectric particles, their $\Im \text{E}$ is expected to be significantly larger than their $\Re \text{E}$, as seen in the calibration measurements presented in Figure \ref{fig:1}b. 
Moreover, a log-normal distribution fit the polystyrene data (Figure 1b) gives a median $\Im \text{E}$ of $1.32\pm0.02\times10^4$~\SI{}{\nm^2} and a median hydrodynamical radius $103\pm$\SI{2}{\nm}.
Using Eq. \eqref{Eq:phase}, these values correspond to a RI of $1.576\pm0.012$ (standard error of mean from log-normal fit), which agrees well with literature values (1.58-1.60) for polystyrene particles\cite{nikolov2000optical,ma2003determination}.

\subsection*{Twilight holographic imaging improves particle contrast while retaining quantitative optical information}
To evaluate the sensitivity and accuracy of twilight holographic microscopy, we measured 150 nm radius silica microspheres (expected RI between 1.43-1.44\cite{van2014refractive,kashkanova2021precision}) suspended in media with varying RI using different water/glycerol mixtures\cite{hoyt1934new}. 
Since the silica microspheres have a RI similar to that of a 70\% glycerol/water mixture, their optical contrast, which is proportional to the difference in RI between the microspheres and the surrounding medium, can be made arbitrarily low by adjusting the glycerol concentration. 
Thus, this experimental system allows us to test the lowest detectable optical signal in our setup.

\begin{figure}
\includegraphics[width=1\textwidth]{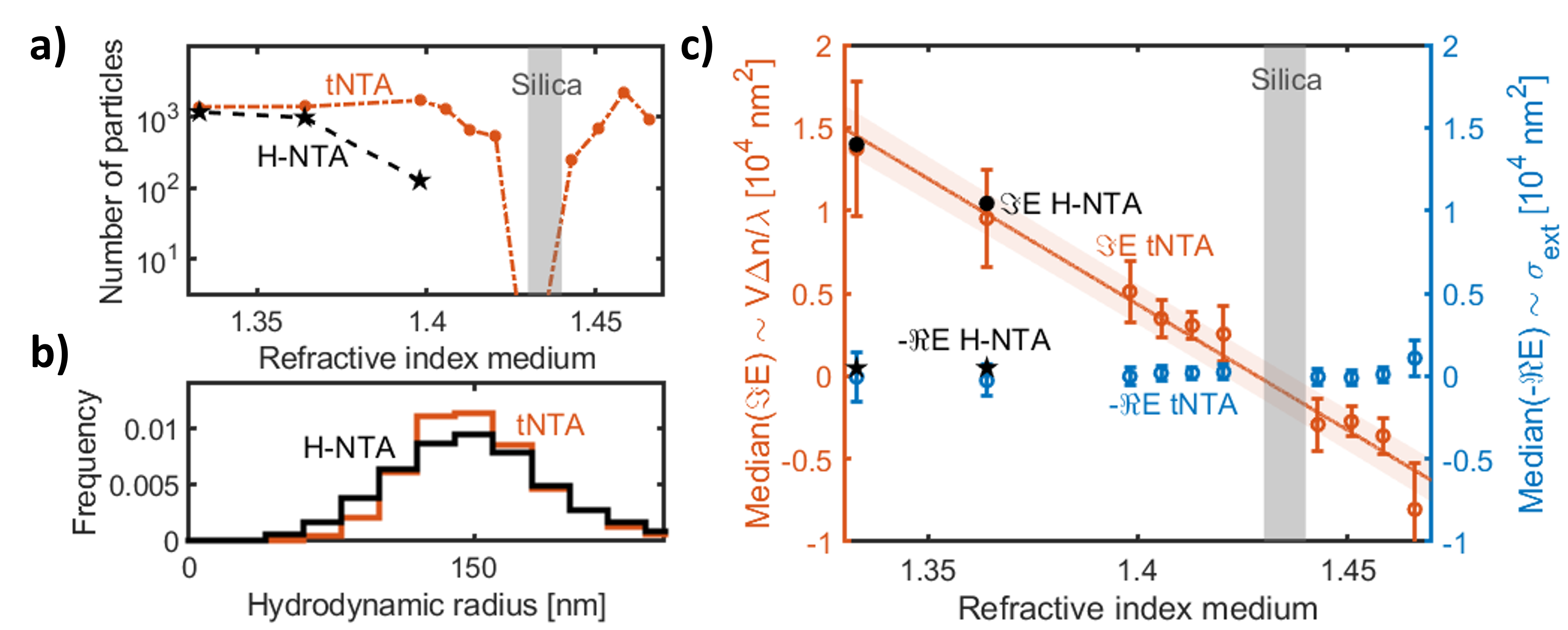} 
\caption{{\bf Quantification of the optical signal as a function of the surrounding refractive index (RI).} {\bf a}, For the two lowest glycerol concentrations, the silica microspheres are observed both with and without the low frequency attenuation filter (LFAF). As the glycerol concentration is gradually increased, the microspheres become only detectable in the presence of the LFAF. The number of observed microspheres is similar for most medium RI values except for the two closest to 1.43, which is expected because the RI difference between the media and the particle approaches zero (expected RI of silica between 1.43-1.44, corresponding to the shaded grey range). {\bf b} Twilight nanoparticle tracking analysis (tNTA) gives the same size as holographic nanoparticle tracking (H-NTA) for silica microspheres in water, indicating that the microsphere population is sampled equally using both techniques. {\bf C} tNTA and H-NTA gives very similar median imaginary signal and median real signal when the particles can be observed with both techniques, further validating that the population is equally sampled both with and without the LFAF. The imaginary part scales approximately linearly with the RI difference. The red line is a linear fit to the imaginary part of tNTA data which intersects the zero-signal level at an RI of $1.428\pm0.008$. The uncertainty of the fit is shown by the blue shaded region. The shaded grey range is the literature values of the RI for silica microspheres of $1.43-1.44$.}
\label{fig:2}
\end{figure}

For pure water and the two lowest glycerol concentrations (RI of $1.364$ and $1.398$) the silica microspheres can be observed both with and without the LFAF, although the number of observed microspheres drops significantly for RI $1.398$ without the LFAF (Figure \ref{fig:2}a). 
When the glycerol concentration is further increased (decreasing the RI difference), the magnitude of the optical signal is reduced and no microspheres are detected without the LFAF; in contrast, in the presence of the LFAF, the number of detected microspheres remains approximately constant up to a bulk RI of $1.420$, at which point the number of detected microspheres rapidly drops to zero. 
Then, the microspheres reappear again at an RI of $1.443$ (Figure \ref{fig:2}a), as expected.
We can conclude that, in the presence of the LFAF, the signal disappears only in a narrow RI difference range of about 0.010-0.015\cite{hoyt1934new,van2014refractive}, indicating that silica microspheres with $\Im \text{E}<0.17\times 10^4$~\SI{}{\nm^2} were not detected, defining the limit of detection of our system. 

For the medium RIs in which the microspheres are detected without the LFAF, the obtained $\Im\text{E}$, $\Re \text{E}$, and hydrodynamic radius coincide when measured with and without the LFAF (Figure \ref{fig:2}b-c).
The measured $\Im\text{E}$ for the silica microspheres using the LFAF scales linearly with the surrounding RI, as predicted from Eq. \eqref{Eq:phase}, while $\Re \text{E}$ remains small throughout the measurement series as expected for dielectric particles (Figure \ref{fig:2}c). 
Also note that the silica microspheres have a lower $\Re\text{E}$ than the polystyrene particles that were used for calibration. 
This is an effect of a reduced backscattering cross section of the silica microspheres due to their weaker RI contrast. 

From the linear fit of the water-glycerol series for the silica microspheres, the $\Im \text{E}$ in water is $(1.45\pm 0.015)\times 10^4$~\SI{}{\nm^2}, which is in good agreement with the expected $\Im \text{E}$ of $1.6\times 10^4$~\SI{}{\nm^2} using a RI for silica of 1.43 in Eq. \eqref{Eq:phase}.
Further, when the dependence of $\Im \text{E}$ on the RI of the surrounding medium is fitted assuming a linear relationship, $\Im \text{E}$ intersects the zero-signal level at an RI of $1.428\pm 0.008$, which coincides with the RI value at which the particles were undetectable (Figure \ref{fig:2}b). 
These results are in agreement with Eq. \eqref{Eq:phase}, as well as with literature values for the RI of silica\cite{van2014refractive}, and indicate that tNTA can be used for particle RI quantification both by combining $\Im \text{E}$ with hydrodynamic radius or by measuring $\Im \text{E}$ at different medium RI.

Notably, the observations closest to a medium RI of 1.43 deviate slightly from the linear fit to $\Im \text{E}$ versus medium RI, while the measurements with an RI difference of at least 0.015 from 1.43 follow the expected trend from Eq. \eqref{Eq:phase}.  
This can be understood from the principal difference between detecting particles and having an accurate ensemble signal, as the latter requires that the particle distribution is sampled in a representative manner.
Thus, the region in (Figure \ref{fig:2}b) within which the signal follows the expected scaling from Eq. \eqref{Eq:phase} defines the limit of full sample characterisation, which corresponds to an absolute particle signal around $0.20\times10^4-0.26\times10^4$~\SI{}{\nm^2}, which is higher than the limit of detection given above ($\sim 0.17\times10^4$~\SI{}{\nm^2}). 
At this point, it is important to note that in order for a particle to be detected, and its complex valued optical signal quantified, it is sufficient that the absolute value, not the real or imaginary part, of the scattered field is above the detection limit. 
This hints at two intriguing possibilities, both of which are explored in the following sections, namely: \textbf{(1)}, the presence of metallic structures bound to dielectric particles with good optical contrast can be detected and quantified at relatively low concentrations, and \textbf{(2)}, the optical signal of particles that are too faint to be detected on their own, can be quantified by tagging them with metallic NPs.


\subsection*{Twilight holographic imaging separates the metal--dielectric signal contributions in particle complexes}

\begin{figure}
\includegraphics[width=1\textwidth]{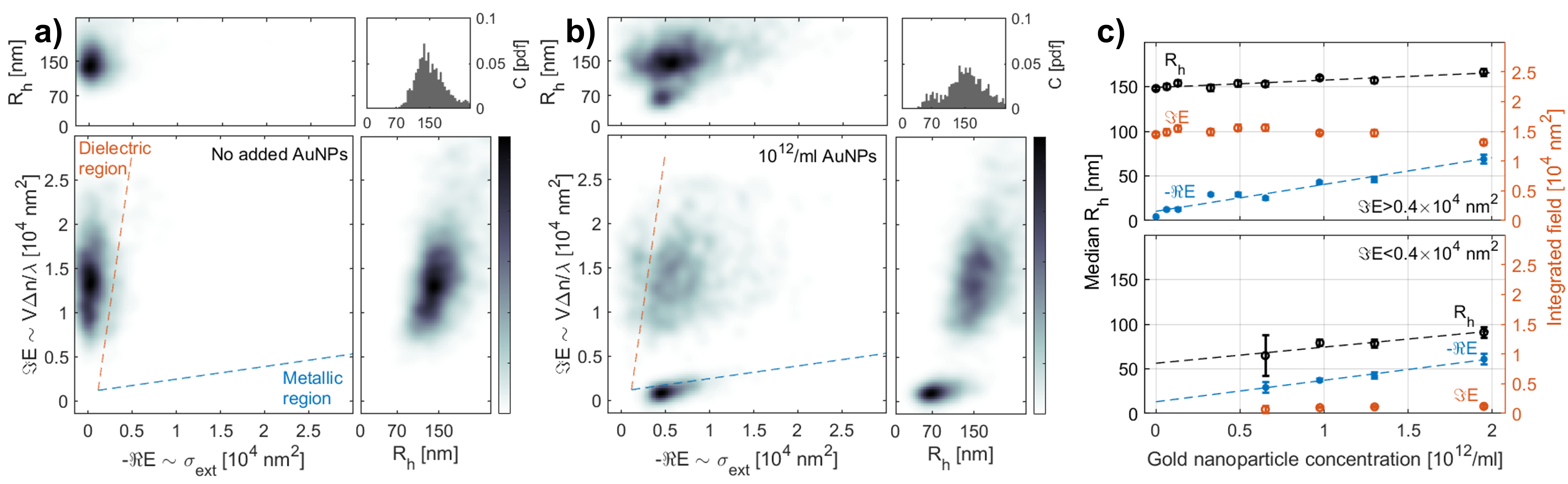} 
\caption{{\bf Salt-induced aggregation of silica microspheres and gold nanoparticles measured using twilight holography.}  {\bf a-b}, The optical particle signal and hydrodynamical radius ({\bf a}) without added gold and ({\bf b}) with $10^{12}$/ml added AuNPs, featuring a shift towards negative-real-part values as the AuNP concentration increases as well as the appearance of a subpopulation with a hydrodynamical radius around 50-100 nm and a high $-\Re \text{E}/\Im\text{E}$ ratio, which agrees with that expected of AuNP aggregates. {\bf c} The median $\Im\text{E}$, $-\Re \text{E}$ and hydrodynamic radius from shifted log-normal fits (Methods, ''Histogram fitting'') as a function of AuNP concentration for the particles with $\Im \text{E}> 0.4\times10^4$~\SI{}{\nm^2} (upper) and $\Im \text{E}< 0.4\times10^4$~\SI{}{\nm^2} (lower). The lines in {\bf c} are linear fits to highlight that the integrated real part ($\Re \text{E}$) and the hydrodynamic radius scales approximately linearly with the AuNPs concentration.}
\label{fig:3}
\end{figure}

To investigate whether the dielectric--metallic material signals of NP complexes can be separated \textit{via} the $\Im\text{E}$ and $\Re \text{E}$ measured using tNTA, as is expected from Eqs. \eqref{Eq:phase}-\eqref{Eq:realpart}, we formed complexes composed of 150 nm radius silica and 5 nm radius AuNPs through salt-induced aggregation at constant silica but varying AuNP concentrations.
In the absence of AuNPs, all detected silica particles have an  $\Im \text{E}$ within 0.5-2.5$\times10^4$~\SI{}{\nm^2}, with a median hydrodynamical radius of $149\pm 2$~\SI{}{\nm} and a $\Re \text{E}$ centred close to 0 (Figure \ref{fig:3}a). 
Salt-induced aggregation of a solution containing both silica and AuNPs causes the signal to shift towards negative $\Re \text{E}$ while $\Im \text{E}$ remains similarly distributed (Figure \ref{fig:3}b); further, a second particle population with a large $-\Re \text{E}/\Im \text{E}$ ratio emerges in the presence of AuNPs. 
Following Eqs. \eqref{Eq:phase}-\eqref{Eq:realpart}, the two populations are attributed to complexes containing both silica and AuNPs, and aggregates of AuNPs, respectively, demonstrating that the optical signals of pure silica, silica/AuNP complexes, and pure AuNPs are clearly distinguishable by their real and imaginary parts.

To make these differences quantitative, we explore the dependence of size, $-\Re \text{E}$, and $\Im \text{E}$ of silica/AuNP complexes on the concentration of AuNPs in the solution during the salt-induced aggregation. 
In order to analyse the different sub-populations, the data is divided into two categories based on the observed $\Im \text{E}$, using a threshold of $0.4\times 10^4$~\SI{}{\nm^2}. 
This threshold corresponds to 25\% of the median $\Im \text{E}$ for the silica particles, above which all silica particles are located in the absence of AuNPs.
The median hydrodynamical radius of the population with $\Im \text{E}>0.4\times 10^4$~nm$^2$ increases from \SI{149}{\nm} to \SI{165}{\nm} with increasing AuNP concentration using a linear fit, which corresponds to an increase in radius similar to the diameter of the AuNPs. 
At the same time, the median $\Im \text{E}$ remains approximately constant for all AuNP concentrations, while the median $-\Re \text{E}$ increases with AuNP concentration.
These observations indicate that the particle populations in all AuNP concentrations consist of individual silica particles with varying amount of bound AuNPs.
In contrast, the median hydrodynamical radius for the particle population below the $\Im \text{E}$ threshold increases linearly from 68 nm to 92 nm with increasing AuNP concentration (Figure \ref{fig:3}c), the $-\Re \text{E}$ increases linearly from 0.44$\times 10^4$~\SI{}{\nm^2} to 0.90$\times 10^4$~\SI{}{\nm^2}, while $\Im \text{E}$ remains close to zero for all concentrations (Figure \ref{fig:3}c).
The observation that $|\Im \text{E}|\ll |\Re \text{E}|$ for this population, strengthens the interpretation that these particles are metallic.

Under the assumption that the absorption cross section of AuNP aggregates is linearly proportional to the number $N_{\rm Au}$ of AuNPs, $\sigma=N_{\rm Au}\sigma_{\rm Au}$ where $\sigma_{\rm Au}=53$~\SI{}{\nm^2} for individual \SI{5}{\nm} radius AuNP \cite{maier2007plasmonics,olmon2012optical}, we find that an aggregate consisting of AuNPs alone must contain approximately 100 NPs to be over the detection limit of $\sim 0.25\times10^4$~\SI{}{\nm^2}. 
A similar calculation yields that aggregates having $-\Re \text{E} = 0.9\times 10^4$~\SI{}{\nm^2} contains approximately 340 AuNPs. 
The expected size of such aggregates can be estimated by viewing them as fractal clusters. 
Salt-induced aggregation of colloids is expected to yield clusters with a fractal dimension of around $2$\cite{lattuada2003hydrodynamic,schaefer1984fractal,fung2019computational}, which indicates that the radius of gyration of the aggregates is expected to scale as $R_{\rm g}\sim N_{\rm Au}^{1/2}$. 
Assuming that the hydrodynamic radius scales similarly as the radius of gyration with the number of monomers, we find that the radius of an aggregate consisting of $100$ and $340$ AuNPs are \SI{46}{\nm} and \SI{84}{\nm}, respectively (Supplementary information, Section 3.2).
This is in good agreement with the obtained size for the population below the $\Im \text{E}$ threshold (Figure \ref{fig:3}c).
Thus, the analysis suggests that salt-induced aggregation of a solution containing both silica particles and AuNPs results in aggregates consisting solely of AuNPs, as well as silica particles (individual or multiple) surrounded by multiple AuNPs, and that aggregates with and without a silica particle can be separated by an $\Im \text{E}$ threshold.  


To specifically investigate how AuNP bound to dielectric particles affect $\Im \text{E}$ of the particle complex, we analyse the particles above the $\Im \text{E}$ threshold.
For that population, the median $\Im \text{E}$ is stable around $1.5\times 10^4$~nm$^2$ within a few percentage points for all AuNP concentrations (Figure \ref{fig:3}c).
Since the collision rate between silica NPs and AuNPs is, to first approximation, proportional to the AuNP concentration\cite{gregory2009monitoring}, the linear change in $\Re \text{E}$ versus AuNP concentration (Figure \ref{fig:3}c) supports that $\sigma$ for the particle complexes scales linearly with AuNP coverage.
This further supports the assumption that the extinction cross section of the particle complexes scales as $\sigma=N_{\rm Au}\sigma_{\rm Au}$, where $N_{\rm Au}$ is the number of AuNPs, and implies that the amount of light reaching each AuNP is undisturbed by the presence of neighbouring AuNPs. 
Under this assumption, the shift in $-\Re \text{E}$ of about $10^4$~nm$^2$ for the highest AuNP concentration corresponds to almost 400 AuNPs. Binding 400 AuNPs to a $150$ nm radius silica particle corresponds to a surface coverage of $\sim$10\% or an inter AuNP distance of $\sim 15$ nm for the AuNPs on the particle surface, assuming that they attach as a monolayer. 
Thus, the data support that the amount of dielectric and metallic material is quantitatively encoded in $\Im\text{E}$ and $\Re \text{E}$, respectively, and that the presence of AuNPs does not significantly affect the dielectric particle signal. 


\subsection*{Twilight holographic imaging can quantify the interaction between tannic-acid-modified gold nanoparticles and herpes simplex viruses}

\begin{figure}
\includegraphics[width=1\textwidth]{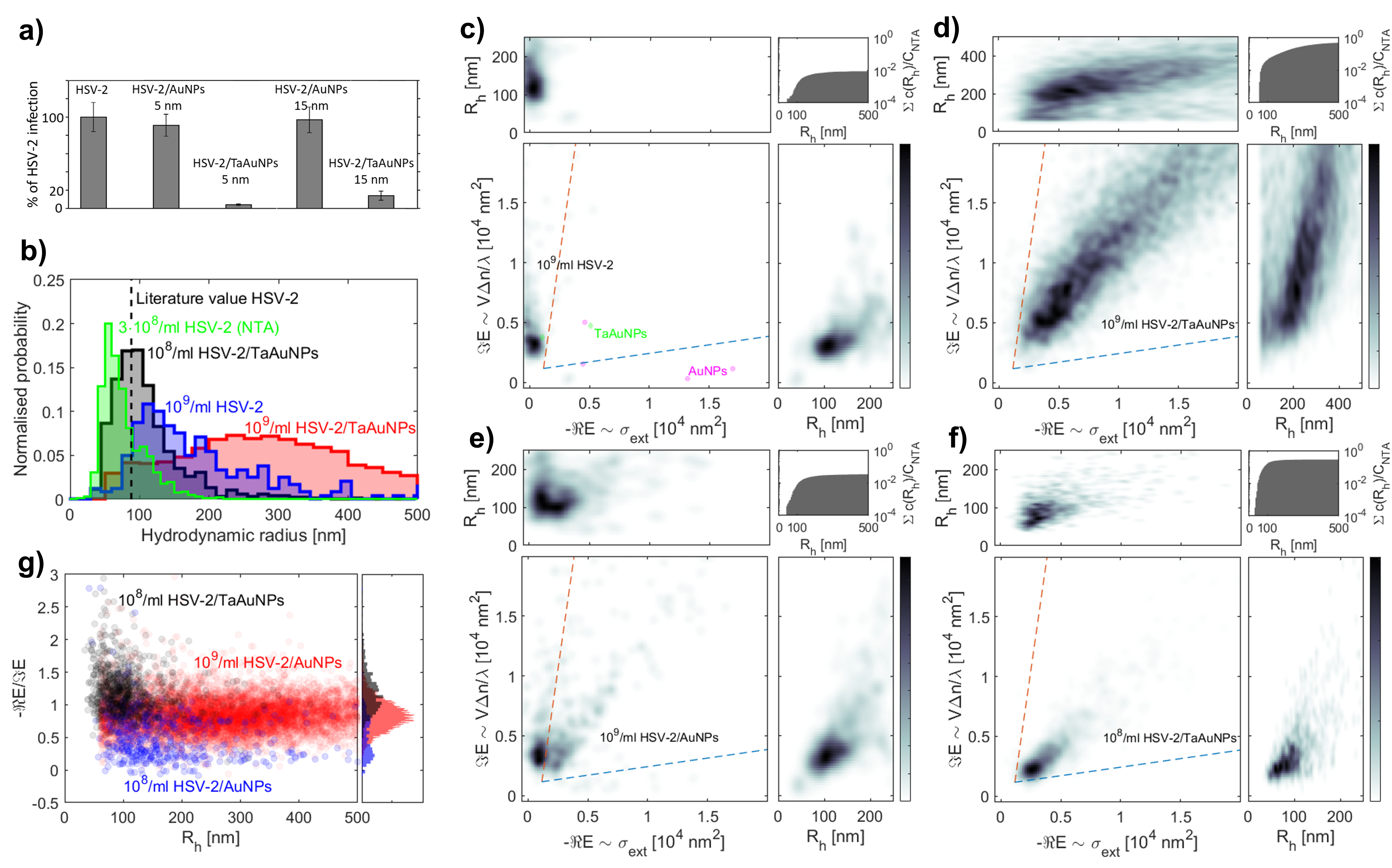} 
\caption{{\bf Viral inhibition assay and twilight measurements of herpes simplex virus type 2 (HSV-2).} {\bf a} HSV-2 titers ($10^7$ PFU/mL) in Vero cells were compared after preincubation with 5/15 nm gold nanoparticles (AuNPs) and tannic-acid-modified AuNPs (TaAuNPs) (\SI{10}{\mu g/ml}, corresponding to $10^{12}$/ml for 5 nm NPs and $3\times 10^{10}$/ml for 15 nm) and expressed as percentage of infected, untreated control (100\%). The results show that TaAuNPs reduces the infectivity of HSV-2. The data are expressed as means from three independent experiments $\pm$ standard error of the mean. {\bf b} Size distributions for HSV-2 measured using both twilight and dark-field NTA when incubated with or without TaAuNPs. The dashed line is the literature value of 88~nm\cite{thorsteinsson2020fret}. {\bf c-f} Twilight measurements of HSV-2, AuNPs and TaAuNPs, either measured separately or HSV-2 mixed with one of the other two. The top-right corner shows the cumulative distribution of detections as a function of size, normalised to the HSV-2 concentration obtained from dark-field NTA. {\bf c} HSV-2, AuNPs and TaAuNPs when measured separately. Only a small fraction of the expected HSV-2 population is detected in tNTA, where the particle sizes are significantly larger than the expected literature value. Only a handful of detections are made in the pure AuNP and TaAuNP samples. {\bf d} $\sim10^9$/ml HSV-2 when mixed with TaAuNPs, for which the measured particle concentration is close to the expected value, the $-\Re \text{E}$ shifts and a broad hydrodynamical radius distribution is induced, which all indicate HSV-2/TaAuNP binding and particle complex aggregation. {\bf e} $\sim10^9$/ml HSV-2 when mixed with AuNPs, where the concentration is more than order of magnitude lower than the expected value and the only a small shift in $-\Re \text{E}$ occur. {\bf f} $\sim10^8$/ml HSV-2 when mixed with TaAuNP, where the concentration is close to that expected, the hydrodynamical radius agree well with the literature value for HSV-2, and the shift in $-\Re \text{E}$ indicate TaAuNP binding. {\bf g} Ratio between $-\Re \text{E}$-$\Im \text{E}$ for the HSV-2/TaAuNP and HSV-2/AuNP data in {\bf d}, {\bf f} and {\bf g}, where a constant ratio for different particles sizes indicates a constant HSV-2/(Ta)AuNP mass ratio for the particle complexes.}
\label{fig:4}
\end{figure}

Tannic-acid-modified silver NPs have been shown to have anti-viral properties for HSV-2\cite{orlowski2014tannic,orlowski2018tannic}, but the mechanism of action is elusive.
Thanks to its capacity for resolving material properties of individual metal/dielectric particles complexes, we hypothesised that twilight holographic imaging might provide insights into the mechanisms of the antiviral activity of tannic-acid-modified metal NPs.
To investigate this possibility, we first studied the antiviral activity of tannic-acid-modified gold NPs. 
We incubated HSV-2 at a concentration of $10^7$ PFU/ml (plaque forming units) with AuNPs and TaAuNPs (with radii of either 5 and 15 nm) at a concentration of \SI{10}{\micro g/ml} ($10^{12}$/ml for 5 nm NPs and $3\times 10^{10}$/ml for 15 nm) and subsequently subjected to Vero cells.
The pre-incubated sample demonstrated a reduction in infectivity of 97 and 88\% for 5 and 15 nm TaAuNPs, respectively, while pure AuNP had insignificant effect compared with the control (Figure \ref{fig:4}a). 
This indicates that tannic-acid-modified NPs have anti-viral properties for HSV-2 also when the core material is changed from silver to gold.

To experimentally investigate the nature of the anti-viral action, we incubated HSV-2 either with AuNPs and TaAuNPs of the same radii and measured the resulting particle complexes using tNTA.
In this way, we were able to detect a fair number of events for the HSV-2 solution at a concentration of $\sim10^{9}$/ml ($\sim 2$ pM) already in the absence of TaAuNPs, but the detection count per unit sample volume is more than a magnitude lower than that obtained in dark-field NTA (Figure \ref{fig:4}c). 
The optical signal of the detections is concentrated along the imaginary axis, as is expected for dielectric particles. 
The hydrodynamic radius and $\Im\text{E}$ of the tNTA detections have median values $R_{\rm h}=152\pm 101$~\SI{}{nm} (STD) and $\Im \text{E}=0.42\times10^4\pm 0.60\times10^4$~\SI{}{nm^2} (STD); thus, both measures have broad distributions whose values are considerably larger than the expected values of $R_{\rm h}=88$ nm \cite{thorsteinsson2020fret} and $\Im \text{E}=0.26\times 10^4$~nm$^2$ for individual HSV-2 viruses assuming a RI of 1.41\cite{ymeti2007fast}. 
Since the signal from a single virus is near the limit of quantification, this observation together with the lower detection count compared to dark-field NTA likely reflects a bias of the tNTA detections toward virus aggregates and other potential co-purified biological materials\cite{gerba2017viral}. 
The presence of larger particles in the HSV-2 sample is further corroborated by observations using conventional dark-field nanoparticle tracking analysis (NTA), which displays a side peak around 110 nm radius (Figure \ref{fig:4}b).

The tNTA measurements of the pure 5 nm radius TaAuNPs and 5 nm AuNPs suspensions display only a handful observations, characterised by large absolute values of $\Re \text{E}$ and low values of $\Im \text{E}$, at a concentration of $\sim3\times10^{12}$/ml (6 nM) (Figure \ref{fig:4}c). 
Since the signal from 5 nm AuNPs is significantly lower than the detection limit, this shows that no major gold NP aggregates are present in the pure AuNP suspensions.

Upon mixing the HSV-2 solution with either TaAuNPs or AuNPs at the same concentration as above, the detection events display significant contributions in both $-\Re \text{E}$ and $\Im\text{E}$ (Figure \ref{fig:4}d-f). 
In contrast, having unbound AuNPs at this concentration does not affect the optical signal of the detected particles as shown by mixing HSV-2 with inert PEG-modified AuNPs (Supplementary information, Figure S11). 
Thus, the shift in the real part observed when mixing HSV-2 with AuNPs and TaAuNPs is indicative of complex formation between gold and dielectric material.

In the HSV-2/TaAuNPs case the detected particle complex concentration is close to half of that obtained using dark-field NTA of pure HSV-2, indicating that the tannic-acid-functionalisation promotes sufficient binding of TaAuNPs to HSV-2 to bring the HSV-2/TaAuNP complexes above the detection limit.
In contrast, for the HSV-2/AuNPs case the relative number of detections remains over one order of magnitude lower than the HSV-2 dark-field NTA concentration, suggesting that the interaction between AuNPs and HSV-2 is significantly weaker than that of TaAuNPs. 
Similar observations of increased number of detection events and a combination of dielectric and metallic signals were made also for 15 nm radius AuNPs and TaAuNPs when mixed with the HSV-2 sample (Supplementary information, Figure S10).
When lowering the virus concentration by one order of magnitude to $\sim10^8$/ml (200 fM) while keeping the TaAuNP concentration the same, the detected particle complex concentration is also similar to that expected from dark-field NTA. 
As the distributions in hydrodynamic radius and optical signal are wide due to sample heterogeneity (Figure \ref{fig:4}), both the mode and median value are used to analyse the data.
The mode and median radius of the $\sim10^8$/ml HSV-2/TaAuNPs complexes obtained from a log-normal fit are \SI{81}{nm} and $91\pm 3$~nm, respectively, values that are close to the expected HSV-2 value (Figure \ref{fig:4}b), while a mode and median radius of \SI{235}{nm} and $282\pm 2$~nm was obtained at the $\sim10^9$/ml HSV concentration.
This suggests that TaAuNPs binds to HSV-2, but that aggregation is induced at the higher HSV-2/TaAuNP ratio. 
The size and optical signal is not greatly affected by further lowering the virus concentration by a factor of 10, and the observation count drops in proportion to the virus concentration (Supplementary information, Figure S12).
Combined with the observation that size and optical signal is unaffected by incubation time (between 10-60 minutes) for the $\sim10^8$/ml HSV-2/TaAuNP sample (Supplementary information, Figure S8), this supports the interpretation that TaAuNP induced aggregation of HSV-2 is negligible at low virus concentrations, indicating that our obtained values at low concentrations relate to particle complexes consisting of a single HSV-2 with multiple bound TaAuNPs.

To investigate whether binding-induced virus rupture underlies the antiviral activity of TaAuNP, we estimate the real part of the RI together with particle radius of the HSV-2/TaAuNP complexes formed at low virus concentrations. 
The mode and median $\Im \text{E}$ are $0.26\times10^4$~\SI{}{nm^2} and $0.32\times10^4$~\SI{}{nm^2}, respectively. Although these values are similar to the limit of quantification introduced above, the \textit{magnitude} of the optical signal is significantly higher than this limit owing to the added signal from the bound TaAuNPs. The mode and median hydrodynamic radius and $\Im \text{E}$ of these complexes (Figure \ref{fig:4}f) together imply an effective modal and median RI of the complexes corresponding to $1.44$ and $1.43$, respectively, which is in good agreement with previous estimates for HSV-2 virus \cite{ymeti2007fast}.
Note that this RI estimate assumes that the TaAuNPs do not affect the particle size. 
As demonstrated in Figure \ref{fig:3}c, binding of AuNPs does in fact slightly increase the hydrodynamic size of the particle complex. 
The true size of the virus in the particle complex is therefore likely slightly smaller than the hydrodynamic size measured in tNTA, and the RI estimate stated above is consequently a slight underestimate of the true RI (by $\sim 0.01-0.02$ RI units).
Moreover, a high amount of bound AuNPs to dielectric particles is expected to slightly decrease $\Im \text{E}$ (Supplementary information, Figure S6), which would lead to a further underestimation of the RI.
If the TaAuNP binding would have induced virus rupture, one should instead expect a loss of virus mass and a measured RI similar to reported values of 1.35 for extracellular vesicles with leaky membranes\cite{rupert2018effective}.
Consequently, TaAuNP binding likely does not induce virus rupture, and HSV-2/TaAuNP complexes appear to consist of intact HSV-2 viruses with TaAuNPs bound on their surfaces. 
This type of binding is further confirmed using cryoTEM imaging, which shows that TaAuNPs bind to the surface of HSV-2 (Supplementary information, Figure S9), and is also consistent with previous studies of the interaction between TaAuNPs and lipid vesicles.\cite{moghadam2012role}.

Having excluded binding induced rupture as the mechanism underlying the antiviral activity of TaAuNP, we next investigate direct particle binding to the virus as a potential model for virus activity inhibition\cite{gurunathan2020antiviral}. 
To quantify the amount of bound TaAuNPs we evaluate the ratio $-\Re \text{E}/\Im \text{E}$ for the complexes formed with AuNPs as well as those formed with TaAuNPs. 
This ratio encodes the ratio of metallic to dielectric mass in the complexes, with a higher ratio corresponding to a higher fraction of metallic mass. 
We find that the HSV-2/TaAuNP complexes have a median $-\Re \text{E}/\Im \text{E}$ ratio of $0.84\pm0.26$ (STD) when using $\sim10^9$/ml HSV-2 and $1.15\pm0.37$ (STD) when using $\sim10^8$/ml HSV-2, which is both higher and less heterogenous than obtained for $0.48\pm1.27$ (STD) for non-functionalized AuNPs using $\sim10^9$/ml HSV-2  (Figure \ref{fig:4}e). 
This ratio can be converted into an estimate of the number of AuNPs bound per virus as well as the surface coverage of NPs on the virus surface.
When using the mode radius and $\Im \text{E}$ for $\sim10^8$/ml HSV-2/TaAuNPs to represent the virus, we find that complexes formed through TaAuNP binding have a significantly higher surface coverage than complexes formed by non-functionalized AuNPs (median $9.5\pm 3.0\%$ (STD) for $\sim10^8$/ml HSV-2/TaAuNPs, median $6.9\pm 2.1\%$ (STD) for $\sim10^9$/ml HSV-2/TaAuNPs, and median $3.9\pm 10.4\%$ (STD) for $\sim10^9$/ml HSV-2/AuNPs).
It should be noted that the detections of complexes formed by non-functionalized AuNPs are likely biased toward complexes with high AuNP coverage, as some AuNP binding is required to bring them over the detection limit. 
Thus, not only does TaAuNP enhance the formation of metal/virus complexes, but the complexes that are formed also have a higher surface coverage of NPs. 
Due to the surface coverage, direct particle binding to the virus therefore emerges as a likely mechanism of action for the antiviral property of TaAuNPs.



\section*{Conclusion}

In this work, we extend the quantitative measurement capabilities of off-axis holography into the nanoparticle regime by combining it with a Low Frequency Attenuation Filter (LFAF), and demonstrate that the effect of the LFAF on the interferometric signal can be quantitatively compensated for in the post-analysis. 
This extension enables multiparametric characterisation of individual suspended complexes formed by binding AuNPs on dielectric particles in terms of their hydrodynamic radius and complex polarizability, which we here represent by the integrated real and imaginary parts of the optical signal by the complexes.
The extinction cross section is dominated by the bound AuNPs, and the dielectric signal is to a first approximation unaffected by the presence of bound AuNPs. 
This permits quantitative measurements of the amount of bound AuNPs and the dielectric mass at the level of individual NP complexes. It was recently demonstrated that the metallic and dielectric contributions of individual nanoparticles can be separated using quadriwave lateral shearing microscopy\cite{khadir2020full,nguyen2023label}, an approach which, if combined with the analytical approach outlined in this paper, has a similar potential for quantification of NP complexes.
Thus, tNTA, or related techniques, will likely find applications in a wide range of biological-metallic hybrid systems, including the interaction between dielectric and metallic NPs, as herein exemplified by quantification of size, refractive index, and gold mass in HSV-2/TaAuNP complexes, as well as in the quantification of the loading in biological-metallic hybrid particles\cite{kulkarni2019fusion,zhigaltsev2022synthesis,wu2022extracellular}.


We have also used the enhanced capabilities of tNTA to investigate the anti-viral properties of TaAuNPs in the context of HSV-2 viruses. 
Since the detected HSV-2/TaAuNPs complexes have a size and RI in agreement with the expected values for HSV-2 while still having a significant TaAuNP coverage, the likely antiviral mechanism of tannic-acid-modified NPs is direct particle binding to the virus, as virus disruption would make the measured size and RI deviate from the expected values for HSV-2. 
As the TaAuNP surface coverage is lower than the jamming limit of $\sim$54\%, the TaAuNP binding appears to be limited by the number of available binding sites for the TaAuNP-virus interaction. 
Considering that tannic acid is known to interact with glycoproteins in the outer envelope of herpes viruses\cite{lin2011hydrolyzable}, it can be speculated that the number of bound TaAuNPs is related to the prevalence of glycoproteins in the outer virus membrane.
However, to get specific information about which molecules that participate in the binding, these types of measurements need to be complemented with suitable competition assays. 

Taken together, our results show that tNTA enables quantitative optical characterisation of heterogeneous suspensions of NP complexes. 
To improve the sample characterisation further, the detection limit can likely be further improved by optimising the suppression of unscattered light as well as by performing the detection and signal quantification using neural networks\cite{midtvedt2021quantitative,midtvedt2022single}. 
Similar to other optical microscopy techniques, twilight off-axis microscopy can also be generalised to investigate particles inside cells, which opens up for analysis of NP-cell interaction to obtain information about both particle uptake as well as their subsequent accumulation or release\cite{wu2022extracellular}.
We therefore anticipate that this type of multiparametric characterisation will find widespread application in any area where heterogeneous particles play an important role, ranging from industrial processes to drug discovery and medical diagnostics.

\section*{Methods}

\subsection*{Particles}
The used commercial particles are 105~nm (modal) radius latex beads (Sigma-Aldrich), 150~nm (modal) radius Silica particles (KISKER BIOTECH GmbH \& Co.KG), 228~nm (modal)  radius (NIST-certified standard
deviation $\pm6.8$ nm) polystyrene (Polysciences), where the sizes were verified using nanoparticle tracking analysis (NanoSight)).
The used gold nanoparticles (AuNPs) were fabricated in-house had a radius of either 5 and 15 nm, where the AuNPs either had no surface functionalisation, tannic acid modification (TaAuNPs) or pegylation (PEGAuNPs).
The herpes simplex virus 2 (HSV-2) suspension was grown in-house.

\subsection*{Gold nanoparticle }\label{Sec:AuSynt}
The AuNP colloids were synthesised in water according to the chemical reduction method, using tetrachloroauric acid (HAuCl$_4$, Sigma-Aldrich, $\geq$ 49\%), tannic acid (C$_{76}$H$_{52}$O$_{46}$, Sigma-Aldrich), and sodium citrate (C$_6$H$_5$Na$_3$O$_7\times 2$H$_2$O, purity 99.0\%, Sigma-Aldrich).
AuNPs with a radius of 5 and 15 nm and weight concentration of Au in colloid equal 100 ppm were synthesised in water by reduction of gold (III) chloride hydrate.
The synthesis procedures are described in detail in previous work\cite{orlowski2018tannic}. 
Briefly, an aqueous solution of gold (III) chloride hydrate was boiled and vigorously stirred under reflux.
Next, a mixture of aqueous solutions of sodium/ammonium citrate and/or tannic acid was added into the solution. 
After the reducing mixture changed colour to red (indicating the formation of AuNPs), the colloids were stirred for an additional 15 min under reflux and cooled down to room temperature. 
Synthesised NPs were characterised using scanning transmission electron microscopy (STEM), dynamic light scattering (DLS) and UV-Vis spectroscopy (Supplementary information, Figure S7). 
STEM measurements were performed using a scanning electron microscope (Nova NanoSEM 450, FEI, accelerating voltage of 30 kV) equipped with a detector for transmitted electron acquisition (STEM II).
The DLS and Zeta potential measurements were carried out using a Nano ZS Zetasizer system (Malvern Instruments) with a He-Ne laser (633 nm) as the light source (scattering angle 173$^{\circ}$, measurement temperature 25$^{\circ}$C; medium viscosity 0.887 mPa$\cdot$s, material RI 1.330). 
DLS measurements were performed in disposable quartz cuvettes and Zeta potential measurements in a disposable folded capillary zeta cell (DTS 1070). 
Next, synthesised and precisely characterised AuNPs were used for TaAuNPs 5 nm and PEGAuNPs 5 nm preparations.

TaAuNPs were prepared by AuNPs incubation with tannic acid, where the procedure is described in detail in previous work\cite{orlowski2018tannic} 
Briefly, an aqueous solution of tannic acid was added (0.009~g, 5\%) to the AuNPs colloid suspension (3~g). 
The final concentration of Ta in colloid was equal to 315 ppm. 
The PEGAuNPs were prepared by incubation of AuNPs with the poly(ethylene glycol) methyl ether thiol (Sigma Aldrich, CH$_3$O(CH$_2$CH$_2$O)nCH$_2$CH$_2$SH, Mw=6000 g/mol) by colloid incubation with an aqueous solution of PEG-ligand (1\%). 
The amount of PEG ligand was equal 3 PEG molecules per 1 nm$^2$ of AuNP surface. 
The presence of a thiol group in the PEG-ligand provides the possibility of direct ligand bonding to the gold surface. The tannic acid molecules present on the surface of NPs desorb and are replaced by PEG molecules which chemically bind to the AuNPs surface.
The effectiveness of functionalization and colloidal stability of AuNPs were characterized with DLS and HR-STEM techniques (Supplementary information, Figure S7).

The used concentration of the used AuNPs, both unfunctionalised, functionalised using tannic acid and functionalised using PEG, was measured using UV-VIS (Agilent Cary 60) using the extinction cross section for AuNPs for that corresponding size.

\subsection*{HSV viruses}\label{Sec:HSVSynt}
HSV-2 (strain 333) was provided by professor Kristina Eriksson from Department of Rheumatology and inflammation Research, University of Gothenburg. 
HSV-2 (strain 333) was grown and titrated in Vero cells (ATCC$^{\circledR}$ CCL-81, Rockville, MD, USA) by the standard plaque assay (PFU/ml)\cite{krzyzowska2011role}.
The particle concentration of the HSV-2 solution was measured using commercial NTA (NanoSight), giving a stock solution of $\sim 3\times10^{10}$/ml.

\subsection*{Anti-viral assay}
To analyse how AuNPs can influence viral infectivity, Vero cells were pre-chilled at 4~$^{\circ}$C for 15 min, then co-treated with NPs and HSV-2 for 1 hour. 
The virus concentration was $10^7$/ml PFU (plaque forming units).
The AuNP and TaAuNP concentrations during the incubation were all \SI{10}{\micro\gram/\milli\liter}, which corresponds $10^{12}$/ml for the 5 nm radius NPs and $3.6\times10^{10}$/ml for the 15 nm NPs.
After this time, virus and NPs were removed and cell monolayers were washed with ice-cold PBS, and further incubated at 37~$^{\circ}$C. 
At 24 h post infection, virus titers were determined by plaque assays (PFU/mL), as described previously\cite{orlowski2014tannic}, which allows the determination the number of effective virions, able to infect cells. 
Pre-incubation of HSV-2 with unmodified and tannic acid-modified AuNPs sized 5 and 15 nm prior to infection showed a tannic acid-dependent virus inactivation by TaAuNPs sized 5 and 15 nm (p $\leq$ 0.001) and no inactivation with unmodified AuNPs (Figure 4a in Main text). 


\subsection*{CryoTEM measurements}
Cryogenic Transmission Electron Microscopy (cryo-TEM) images were obtained using a Tecnai F20 X TWIN microscope (FEI Company, Hillsboro, Oregon, USA) equipped with a field emission gun, operating at an acceleration voltage of 200 kV. 
Images were recorded on a Gatan Rio 16 CMOS 4k camera (Gatan Inc., Pleasanton, California, USA) and processed with Gatan Microscopy Suite (GMS) software (Gatan Inc., Pleasanton, California, USA). 
Specimen preparation was done by vitrification of the aqueous solutions on grids with holey carbon film (Quantifoil R 2/2; Quantifoil Micro Tools GmbH, Großlöbichau, Germany). 
Prior to use, the grids were activated for 15 seconds in oxygen plasma using a Femto plasma cleaner (Diener Electronic, Ebhausen, Germany).
Cryo-samples were prepared by mixing the viral stock of HSV-2 with $20~\mu$g/mL TaAuNPs ($\sim 2\times 10^{12}$/ml) at 1:1, resulting in HSV-2/TaAuNP ratio of around 1-100, followed by 5 minutes of incubation at room temperature. 
Next, a droplet (3~$\mu$L) of the suspension was added onto the TEM grid, blotted with filter paper and immediate frozen in liquid ethane using a fully automated blotting device Vitrobot Mark IV (Thermo Fisher Scientific, Waltham, Massachusetts, USA).
After preparation, the vitrified specimens were kept under liquid nitrogen until they were inserted into a cryo-TEM-holder Gatan 626 (Gatan Inc., Pleasanton, USA) and analysed in the TEM at -178~$^{\circ}$C. 
The Cryo-TEM measurements were performed under a contractual service agreement with CMPW PAN in Zabrze, Poland.

\subsection*{The microscope setup}
The twilight off-axis holography microscope consists of an off-axis holographic microscope enhanced with a low frequency attenuation filter (LFAF) (Figure \ref{fig:1}). 
The off-axis holographic microscope is custom-built and consists of a continuous 100 mW $\lambda$ = 532 nm diode-pumped solid-state (DPSS) laser (Roithner Lasertechnik GmbH), a 40x oil-immersion objective with a numerical aperture (NA) of 1.3 (Olympus) and a camera (AlliedVision, ProSilica GX1920). 
The laser beam is split into two beams, where the intensities of the two beams are controlled by a $\lambda$/4 plate and a polarisation-dependent beam splitter.
After the polarising beam splitter, these beams are focused into two optical fibers which guide the light to the sample and reference beam path.  
One of the optical fibers is positioned just above the sample, where the laser beam is expanded to an area of approximately 1 cm$^2$ before impinging on the sample. 
The illumination intensity is around a few tens of mW per square centimeter, which is several orders of magnitude lower than what is required to induce significant laser-induced heating \cite{taylor2019interferometric}.
The second laser beam is expanded to a few centimetres in diameter before both beams are recombined before the camera at a slight offset angle using a beam splitter, where the offset angle separates the interference terms when the recorded image is Fourier transformed (Supplementary information, Figure S1).
The effective pixel size of the setup is obtained by measuring the diffusivity of NIST-certified polystyrene particles (Supplementary information, Figure S5).

\subsection*{Experimental conditions}
Each measurement is performed using a sample volume of around $20~\mu$l placed in the inlet of a straight microfluidic chip (20 $\mu$m high and 800 $\mu$m wide, ChipShop).
The flow through the microfluidic channel is gravitationally driven and controlled by the volumes in the inlet and outlet, except when the solution contained 80~wt\% or more glycerol where a syringe pump was used. 
To minimize the risk of particles sticking to the wall of the microfluidic channel, each measurement series starts with incubating the channel using a 5 mg/ml BSA solution (Sigma-Aldrich) during at least five minutes.
The frame rate is 41 frames per second, the exposure time is between 1-2 ms and each recorded video typically consists of 2000 frames. 
The temperature of the room is 22 $^{\circ}$C.
For all included data, a minimum track length of 20 frames was used, while the flow speed of the particles during the measurements was adjusted such that the average track length is around 50-100 frames.

\subsection*{Obtaining the conversion factor}
The conversion factor between the tNTA signal and the signal without LFAF was obtained by measuring a sample of 105~nm median radius latex beads both with and without the LFAF.
The particle suspension was diluted in MilliQ until approximately 100 particles were in the field of view.
The change in beam intensity when the LFAF was removed was compensated by adjusting the coupling into the optical fibre for the beam that passed though the sample.
The search for that factor was performed by using the function {\tt lsqnonlin} in Matlab.
The residuals of the fit were used to estimate the uncertainty.

\subsection*{The low frequency attenuation filter}
The gold disc was made in the cleanroom at Chalmers University of Technology.
A circular cover-slip underwent a dehydration bake consisting of 5 minutes on a hotplate at 190-200~$^{\circ}$C. 
The cover-slip was spin coated with LOR 1A at 3000 rpm for 45 seconds, followed by prebaking on a hotplate (5 minutes at 190-200~$^{\circ}$C). 
It was thereafter spin coated with S1813, 3000 rpm for 45 seconds and prebaked on a hotplate (5 minutes at 110~$^{\circ}$C). 
The sample was then exposed using a laserwriter and developed in MFCD 26 for 60 s, followed by rinsing in deionized water and descum using oxygen plasma. 
The metal was thereafter deposited using e-beam evaporation, first 3 nm titanium and then 55 nm gold, followed by lift-off in 1165 remover for approximately 100 minutes and subsequent rinsing with isopropanol and deionized water. 

\subsection*{Silica-glycerol measurements}
Each particle solution had a combined weight of 2 grams.
It contained 10 $\mu$l 150~nm radius silica solution (KISKER BIOTECH GmbH \& Co.KG) that previously has been 100 times diluted from the stock solution to keep the particle concentration the same.
The glycerol (Sigma-Aldrich) was weighed and corresponded to 0, 25, 50, 55, 60, 65, 70, 75, 80, 85, 90 and 95 wt\%.
MilliQ water was added until the combined mass of the solution reached 2 grams. 
The RI of the water-glycerol mixtures was assumed to be 1.333, 1.364, 1.398, 1.406, 1.420, 1.428, 1.435, 1.443, 1.451, 1.458 and 1.466\cite{hoyt1934new}.

As the flow speed can be slightly different between measurements, the number of detected particles in the data set was normalised by taking the ratio between the median flow speed and the flow speed for that particular measurement to enable number comparison between measurements. 
The flow speed was estimated from the average particle speed for the included particles in the analysis.

\subsection*{Silica-gold measurements}
The Silica particles are 150~nm median radius (KISKER BIOTECH GmbH \& Co.KG).
The stock solution of silica particles is 50 mg/ml, which corresponds to $\sim1.3\times10^{12}$ particles per ml.
In each experiment, 50~$\mu$l 1-1000 diluted Silica solution was mixed with unfunctionalised AuNPs with a varying volume between 0-30~$\mu$l, with a stock concentration of  $6.5\times10^{12}$ ml$^{-1}$.
The used AuNP volumes were 0, 1, 2, 5, 7.5, 10, 15, 20, 30~$\mu$l, making that concentration in the 100~$\mu$l solution during the aggregation ranged from $0-2\times10^{12}$/ml.
MilliQ water was added so that the combined volume was 80~$\mu$l.
To that 20~$\mu$l of saturated NaCl solution is added and mixed, and after 10 seconds 1~ml of  MilliQ water is added to slow down the aggregation.
The mixture was then directly injected into the microfluidic chip, where the time from dilution to measurement was 2-3 minutes.
The silica-to-gold ratio for the highest AuNP concentration is around 1 to 2000.

\subsection*{The HSV-TaAuNP measurements}
All samples were diluted in filtered saline NaCl solution. 
Unless otherwise stated, the incubation time with the HSV-2/different gold NP was 1 hour before the measurement.
All AuNP/TaAuNP/PEGAuNP suspensions were diluted a factor of two from stock solution in the HSV-AuNP mixture.
The HSV-2 suspension was diluted either 30, 300 or 3000 times from the stock particle concentration of $\sim 3\times 10^{10}$ ml$^{-1}$ in the HSV-AuNP mixture.

\subsection*{Data analysis}
All data analysis was performed using MATLAB (Mathworks Inc.).
The data analysis pipeline consists of: 1) obtaining the complex optical signal using standard methods for off-axis holography, 2) normalising the optical signal such that the background signal is equal to one, 3) subtracting the background signal, 4) identifying the particles and re-propagating such that each particle is in focus, 5) repeating for all frames and track the particles, and 6) once a traces is complete, compensating the effect of the LFAF in the post-acquisition. 

The interference patterns, or holograms, were analysed using the software MATLAB (Mathworks Inc.) to extract the amplitude and phase maps using standard methods\cite{kim2010principles}.
In brief, due to the off-axis configuration of the setup, the Fourier transform of the interference pattern contains two off-center peaks, which consists of the complex-valued object signal multiplied by a plane wave ($\exp{\pm i\vec{k}\cdot \vec{R}}$), in addition to the central peak which corresponds to the non-interferometric intensities. 
The distance between the off-center peaks and the central peak is controlled by the relative angle between the object and reference beam. 
In order to isolate the object signal, we numerically shifted one of the off-center peaks to the center of the Fourier spectrum (multiplying the recorded image by $\exp{\pm i\delta \vec{k}\cdot \vec{R}}$) and applied a low-pass filter. 
The magnitude and phase of the isolated signal correspond to the amplitude and phase of the optical signal recorded by the camera. 
The obtained signal is slightly phase distorted due to optical aberration in the optical pathway. 
This was corrected by fitting the phase of the field to a fourth-order polynomial, which was subsequently subtracted from the phase to obtain a phase map. 
The optical signal was thereafter normalised such that the background was centred around 1.
Using normalised field-images, the background signal was subtracted using a linear combination of frames that was within 50 frames of the current frame but not the closest $\pm 3$ frames (to avoid self-subtraction). 
The choice of background frames was based on a greedy search algorithm where a list of potential candidate field was iterated through sequentially, and an element in this list was included in the background subtraction if its inclusion lowered the total power of the Fourier transform of the background subtracted field.

In-plane subpixel localisation and linking between subsequent observations is 
based on the same principles as described in \cite{midtvedt2019size}.
In short, a depth stack of complex-valued optical images was projected onto a 2D plane, where in-plane subpixel localisation of detected local extrema of the signal amplitude was performed using the radial center method\cite{parthasarathy2012rapid}.
Depth localisation is based on minimizing the standard deviation of the Fourier image of single particle detection, which assumes point-like particles\cite{midtvedt2019size} (Supplementary information, Section 2).

\subsection*{Histogram fitting}
Log-normal distributions cannot take negative values, which here well describes for both the hydrodynamic radius and $\Im \text{E}$ when the difference in RI to media is positive, but not for $-\Re \text{E}$ as it can take negative values due to image noise. 
To have similar fits for all data shifted log-normal fits was used.
Specifically, the $-\Re \text{E}$ is shifted such that it lowest value is 0.1, where the shift is subtracted after the fit to quantify the values.

\subsection*{Size estimations from particle tracks}
The radius of the tracked particles was estimated using the Stokes-Einstein relation (at temperature 22~$^{\circ}$C).
To accurately relate mean squared displacement to diffusivity and hydrodynamical radius, the contribution from the position uncertainty needs to be removed\cite{vestergaard2014optimal}.
To estimate the effect of the position uncertainty, the average of subsequent steps was used\cite{vestergaard2014optimal}, where the median uncertainty of all included particles was used to estimate the position uncertainty for all particles in the same measurement.

\section{Acknowledgements}
This research was funded by the Swedish research council, grant number 2019-05071, the Knut and Alice Wallenberg Foundation grant number 2019-0577, the Polish National Science Centre grant number 2018/31/B/NZ6/02606 and Chalmers Area of Advance Nano. 
Myfab is acknowledged for support and for access to the nanofabrication laboratories at Chalmers.
We also thank Dr. Björn Agnarsson for fabricating the LFAF.

\section{Author contributions}
EO, FH and DM conceived the method. EO designed, implemented and tested the method. 
BM, AR, FE, FH and DM contributed to the development of the method.
EO and AR collected the twilight data, using an experimental setup and software developed by EO, BM, DM, and FH. 
KRS, JG and MK fabricated the used gold nanoparticles, surface functionalised the particles, characterised the particles and isolated the HSV-2 solution.
GV, FH and DM supervised the work. 
EO and DM drafted the paper.
EO, MK, GV, FH and DM drafted the illustrations. 
All authors revised the paper.

\section{Competing interests}
The authors declare the following competing financial interest(s): E.O., B.M., F.E. and D.M. own shares in a private company (Holtra) that holds IP related to the twilight microscopy method explored in this work.




\bibliography{references}

\end{document}